\documentclass[aps,prb,superscriptaddress,twocolumn,showpacs,amsmath,amssymb]{revtex4}

\usepackage{color}
\usepackage{graphicx}
\usepackage{bm}

\begin{document}

\title{Electron-hole puddles in the absence of charged impurities}
\author{Marco Gibertini}
\affiliation{NEST, Istituto Nanoscienze-CNR and Scuola Normale Superiore, I-56126 Pisa,
Italy}
\author{Andrea Tomadin}
\affiliation{NEST, Istituto Nanoscienze-CNR and Scuola Normale Superiore, I-56126 Pisa,
Italy}
\author{Francisco Guinea}
\affiliation{Instituto de Ciencia de Materiales de Madrid (CSIC), Sor Juana In\'es de la Cruz 3, E-28049 Madrid, Spain}
\author{Mikhail I. Katsnelson}
\affiliation{Radboud University Nijmegen, Institute for Molecules and Materials, NL-6525 AJ Nijmegen, The
Netherlands}
\author{Marco Polini}
\email{m.polini@sns.it}
\homepage{http://qti.sns.it}
\affiliation{NEST, Istituto Nanoscienze-CNR and Scuola Normale Superiore, I-56126 Pisa,
Italy}

\date{\today}

\begin{abstract}
It is widely believed that carrier-density inhomogeneities (``electron-hole puddles") in single-layer graphene on a substrate like quartz are due to charged impurities located close to the graphene sheet. In this Rapid Communication we demonstrate by using a Kohn-Sham-Dirac density-functional scheme that corrugations in a {\it real} sample are sufficient to determine electron-hole puddles on length scales that are larger than the spatial resolution of state-of-the-art scanning tunneling microscopy.
\end{abstract}

\maketitle

{\it Introduction. ---} Graphene, a single layer of carbon atoms arranged in a honeycomb geometry, is a two-dimensional (2D) system whose carriers are subject to a large number of scattering mechanisms affecting its transport properties in a number of intriguing ways~\cite{reviews,peres_rmp_2010,dassarma_rmp_2011}. When a graphene sample produced by mechanical exfoliation is deposited on a substrate like ${\rm SiO}_2$, it displays a maximum mobility $\approx 1.0 \times 10^4 - 1.5 \times 10^4~{\rm cm}^2/({\rm V s})$. The main scattering mechanism limiting the mobility of such samples is to date still unclear and the subject of a very intense debate~\cite{peres_rmp_2010,dassarma_rmp_2011}.

Martin {\it et al.}~\cite{martin_natphys_2008} were the first to demonstrate by means of a single-electron transistor (SET) that close to the charge neutrality point the carrier density distribution in a graphene sheet is highly inhomogeneous. Disorder-induced potential fluctuations break up the electron liquid into ``electron-hole puddles". These findings have been subsequently confirmed by other groups~\cite{zhang_nature_2009,deshpande_prb_2009,teague_nanolett_2009} by means of scanning tunneling spectroscopy (STS). The typical STS spatial resolution is roughly two orders of magnitude higher than that of the SET employed in Ref.~\onlinecite{martin_natphys_2008} ($\ell_{\rm SET} \approx 150~{\rm nm}$).

\begin{figure}[t]
\begin{center}
\includegraphics[width=1.00\linewidth]{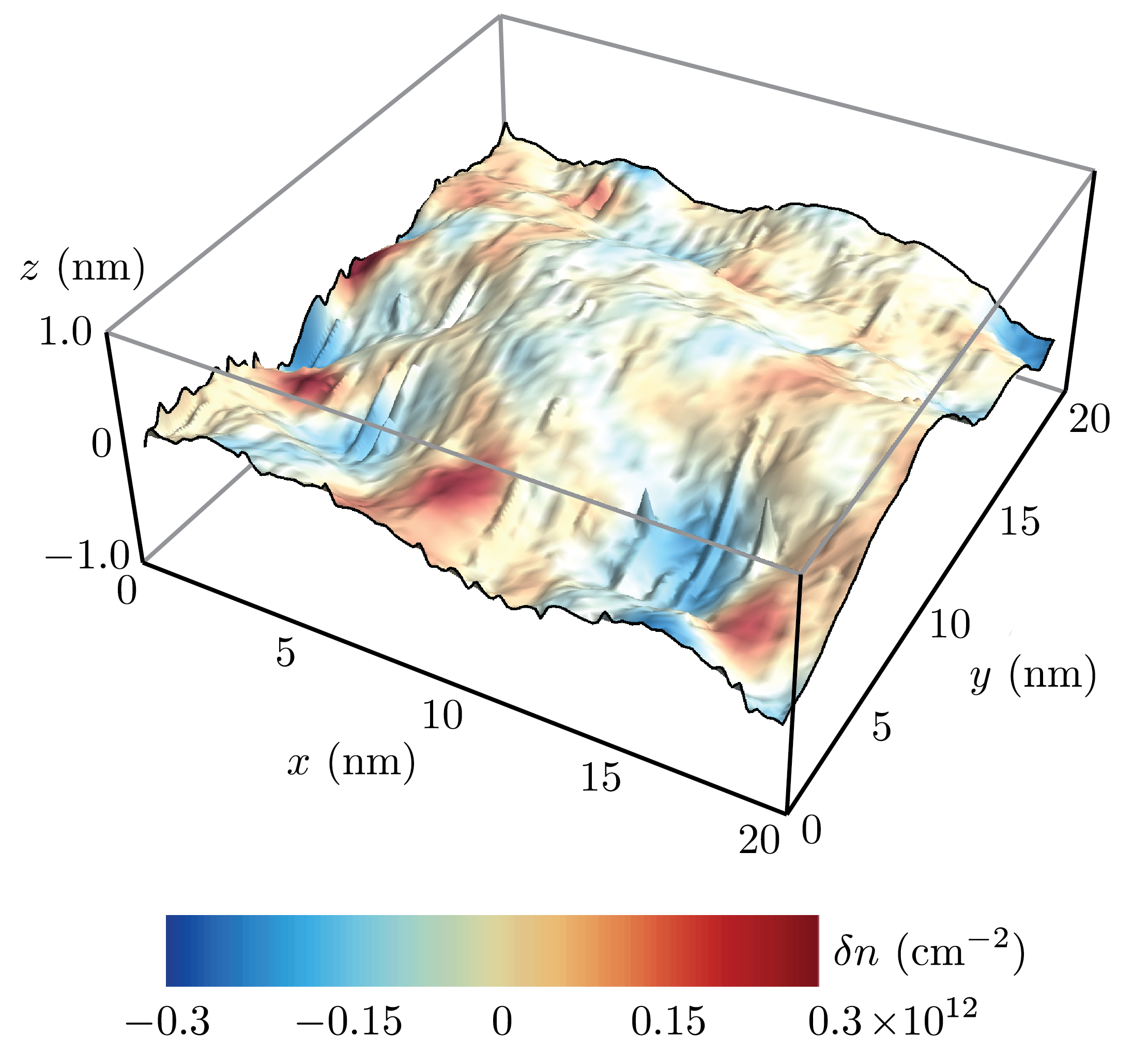}
\caption{(Color online) Three-dimensional plot of the corrugated graphene sample studied in this work (experimental data are a courtesy of V. Geringer~\cite{geringer_prl_2009}). The color-coding of the surface labels the local value of the induced carrier density $\delta n({\bm r})$ as calculated from the Kohn-Sham-Dirac self-consistent theory, Eqs.~(\ref{eq:ksd})-(\ref{eq:density}). The data in this figure have been obtained by setting $g_1 = 3~{\rm eV}$, $\alpha_{\rm ee} = 0.9$, and ${\bar n}_{\rm c} \approx 2.5\times 10^{11}~{\rm cm}^{-2}$ (see text).\label{fig:one}}
\end{center}
\end{figure}

Due to the linear dependence of conductivity on carrier density~\cite{reviews}, charged impurities located near the graphene sheet have been early on recognized as important actors~\cite{chargedimpurities} and
invoked~\cite{marylandpuddles} to predict electron-hole puddles. Quantitative theories of carrier-density inhomogeneities taking into account many-body effects have also been put forward~\cite{rossi_prl_2008,polini_prb_2008}. Despite other alternatives such as frozen ripples~\cite{katsnelson_ptrsA_2008} and resonant scatterers~\cite{katsnelson_ptrsA_2008,RSs} have been proposed, long-range Coulomb disorder is currently the most ``popular" candidate for the {\it main} scattering mechanism limiting mobility in samples on a substrate~\cite{dassarma_rmp_2011}.

Charged-impurity scattering as the main mechanism of disorder has faced, however, severe experimental (and theoretical) difficulties. Ponomarenko {\it et al.}~\cite{ponomarenko_prl_2009} have studied exfoliated samples deposited on various substrates and found a rather weak dependence of the mobility on the type of substrate. In particular, the authors of Ref.~\onlinecite{ponomarenko_prl_2009} have studied transport in flakes embedded in media with high dielectric constants, such as glycerol, ethanol, and water, and measured only a small increase in the mobility (at temperatures above the freezing temperature of these substances). Couto {\it et al.}~\cite{couto_arXiv_2011} have recently reported on low-temperature transport properties of graphene on ${\rm SrTiO}_3$, a well-known insulator with a dielectric constant varying (with temperature) in the range $3 \times 10^2 \lesssim \epsilon_{\rm sub} \lesssim 5 \times 10^3$. The authors of this work have clearly demonstrated that i) neither the carrier mobility nor the amplitude of the carrier-density fluctuations $\delta n$ are affected by the large change in the dielectric constant of the substrate and ii) these quantities are practically identical to those measured in a typical graphene sheet on ${\rm SiO}_2$.

From the theoretical point of view, we will show elsewhere~\cite{gibertini_tobesubmitted} that charged impurities randomly located on a plane (parallel to and) at an average distance $d \approx 1~{\rm nm}$ from the graphene sheet~\cite{distance} create extremely sharp features in the carrier-density spatial profile, in stark contrast with the smooth inhomogeneities measured using STS~\cite{zhang_nature_2009,deshpande_prb_2009}. Moreover, the Dirac-point mapping procedure exploited in Refs.~\onlinecite{zhang_nature_2009} and~\onlinecite{deshpande_prb_2009} fails to yield trustable results for the reconstructed carrier density at distances $d \lesssim  2~{\rm nm}$~\cite{gibertini_tobesubmitted}.

Motivated by this large body of literature, in this Rapid Communication we demonstrate that, contrary to the common wisdom~\cite{peres_rmp_2010,dassarma_rmp_2011,marylandpuddles}, charged impurities are not a necessary ingredient for the existence of electron-hole puddles close to charge neutrality. We establish indeed that smooth electron-hole puddles emerge also in the presence of scalar and vector potentials induced by corrugations only. Carrier density inhomogeneities stemming from ripples and corrugations have already been studied by a few authors~\cite{gibertini_prb_2010, partoviazar_prb_2011}. These studies, however, have focussed on artificial samples whose ripples have been calculated by Monte Carlo or molecular dynamics simulations. The key added value of the present work is twofold: i) we study a {\it real} sample using STS experimental data~\cite{geringer_prl_2009} for the height fluctuations of a graphene sheet on ${\rm SiO}_2$; and ii) we present an approximate theory that allows to calculate corrugation-induced scalar and vector potentials from the knowledge of the STS height-fluctuation maps.

{\it From height fluctuations to scalar and vector potentials. ---} We analyze the $20~{\rm nm} \times 20~{\rm nm}$ corrugated graphene sample shown in Fig.~\ref{fig:one}. The modulations in the height are defined by a height-corrugation profile $h({\bm r})$,
where ${\bm r} = (x,y)$ is a 2D vector. The function $h({\bm r})$ is known experimentally~\cite{geringer_prl_2009}. Modulations in the height lead to stresses and to effective scalar and gauge potentials which couple to the orbital degrees of freedom of the electron gas in the sheet thereby changing the electronic spectrum~\cite{vozmediano_pr_2010}.  In what follows we lay down an approximate theory that allows us to calculate corrugation-induced scalar and vector potentials from the knowledge of the map ${\bm r} \mapsto h({\bm r})$.

We introduce the deformation tensor~\cite{vozmediano_pr_2010,Landau7,guinea_prb_2008,gazit_prb_2009}  $u_{ij} = u_{ij}({\bm r})$ as
\begin{equation}\label{eq:deformationtensor}
u_{ij} = \frac{1}{2} (\partial_j u_i + \partial_i u_j + \partial_i h \partial_j h)~,
\end{equation}
where $u_i$ with $i = x,y$ are the Cartesian components of the 2D displacement vector
${\bm u} =  (u_x, u_y)$ and $\partial_x$ ($\partial_y$) is a shorthand for $\partial/\partial x$ ($\partial/\partial y$).
In writing Eq.~(\ref{eq:deformationtensor}) we have neglected two non-linear terms, {\it i.e.} $(\partial_i u_x)(\partial_j u_x)$ and $(\partial_i u_y)(\partial_j u_y)$, which are at least one order of magnitude smaller that the other terms. The only non-linear contribution to $u_{ij}$ we have retained is the last term of Eq.~(\ref{eq:deformationtensor}), which is of the same order of magnitude of the first two terms in the same equation.

The free-energy of the lattice in the presence of deformations can be written as $E[{\bm u}, h] = \int d^2{\bm r}~{\cal E}_{\rm el}[{\bm u}({\bm r}), h({\bm r})]$ where the elastic free-energy density per unit area ${\cal E}_{\rm el}$ is given by~\cite{vozmediano_pr_2010,Landau7,guinea_prb_2008,gazit_prb_2009}
\begin{eqnarray}\label{eq:elastic}
{\cal E}_{\rm el} = \frac{\kappa}{2} \left[\nabla^2_{\bm r}h({\bm r})\right]^2 + \frac{\lambda}{2}\left[\sum_{i}u_{ii}({\bm r})\right]^2 + \mu \sum_{i,k} u^2_{ik}({\bm r})~.
\end{eqnarray}
Here $\kappa \approx 1~{\rm eV}$ is the bending rigidity and $\lambda = 2.57~{\rm eV}~{\rm \AA}^{-2}$ and $\mu = 9.95~{\rm eV}~{\rm \AA}^{-2}$
are the Lam\'{e} constants of graphene~\cite{zakharchenko_prl_2009} at a temperature $T=300~{\rm K}$ ($\mu$ has the physical significance of shear modulus).
In what follows we neglect the first term in Eq.~(\ref{eq:elastic}) since this is important only at length scales $\ell \lesssim (h/|{\bm u}|)(\kappa/\lambda)^{1/2}
\approx 1~{\rm nm}$ (estimating $h \approx 1~{\rm nm}$ and $|{\bm u}| \approx 0.5~{\rm \AA}$).

The equilibrium condition in the absence of external forces reads $\sum_{k} \partial_k \sigma_{ik} = 0$, where $\sigma_{ik} = \delta E[{\bm u}, h]/\delta u_{ik} = \lambda~\delta_{ik} \sum_j u_{jj}({\bm r}) +2 \mu~u_{ik}({\bm r})$ is the stress tensor~\cite{Landau7}. Solving the two equilibrium equations for $i=x,y$ allows us to calculate the induced in-plane displacements ${\bm u}({\bm r})$ and the deformation tensor $u_{ij}({\bm r})$. In Fourier transform with respect to ${\bm r}$ we find:
\begin{equation}\label{eq:strains}
u_{ij}({\bm q}) =\left[\frac{( \lambda + \mu )}{( \lambda + 2 \mu )}\frac{q_i q_j}{|{\bm q}|^4} - \frac{ \delta_{ij}}{2|{\bm q}|^2}\right]~{\cal F}({\bm q})~,
\end{equation}
where ${\cal F}({\bm q}) \equiv \sum_{i,k} q_i q_k f_{ik}({\bm q}) - |{\bm q}|^2\sum_i f_{ii}({\bm q}) =
2 q_x q_y f_{xy}({\bm q}) - q_y^2 f_{xx}({\bm q}) - q_x^2 f_{yy}({\bm q})$ and $f_{ij}({\bm q})$ is the Fourier transform of the
tensor field $f_{ij} ({\bm r}) = \partial_i h({\bm r}) \partial_j h({\bm r})$.

Scalar $V_1$ and vector $V_2 = A_x - i A_y$ potentials can be easily calculated from the following relations~\cite{manes_prb_2007}
$V_1 = g_1 (u_{xx} + u_{yy})$ and $V_2 = g_2 (u_{xx} - u_{yy} + 2 i u_{xy})$, where $g_1$ and $g_2$ are two coupling constants. Using Eq.~(\ref{eq:strains}) we find
\begin{eqnarray}\label{eq:potentialsfromripples}
\left\{
\begin{array}{l}
{\displaystyle V_1({\bm q})= - g_1 \frac{\mu}{\lambda + 2 \mu}~\frac{q_x^2 + q_y^2}{|{\bm q}|^4}{\cal F}({\bm q})} \vspace{0.2 cm}\\
{\displaystyle A_x ({\bm q}) = g_2 \frac{\lambda + \mu}{\lambda
+ 2 \mu}~\frac{q_x^2 - q_y^2}{|{\bm q}|^4}{\cal F}({\bm q})} \vspace{0.2 cm}\\
{\displaystyle A_y ({\bm q}) = - 2 g_2\frac{\lambda + \mu}{\lambda
+ 2 \mu}~\frac{q_x q_y}{|{\bm q}|^4}{\cal F}({\bm q})}
\end{array}
\right.~.
\end{eqnarray}
For the coupling constant $g_1$ we use the values $g_1 = 3~{\rm eV}$ and $g_1 = 20~{\rm eV}$~\cite{g-one-discussion}, while
$g_2 = 3 c \beta\gamma_0/4$, where $\beta = - \partial  \log{(\gamma_0)}/\partial{\log (a_0)} \approx 2$, $\gamma_0 \approx 2.7~{\rm eV}$ is the nearest-neighbour hopping parameter, 
$a_0 \approx 1.42$~\AA~is the carbon-carbon distance, and $c  \equiv \mu/(B \sqrt{2})$. For the bulk modulus ($B = \lambda + \mu$) we use $B= 12.52~{\rm eV}$~\AA$^{-2}$ at $T = 300~{\rm K}$~\cite{zakharchenko_prl_2009}.
We thus find that $c \approx 0.56$ at this temperature.

\begin{figure*}
\begin{center}
\begin{tabular}{c c c}
\includegraphics[width=0.33\linewidth]{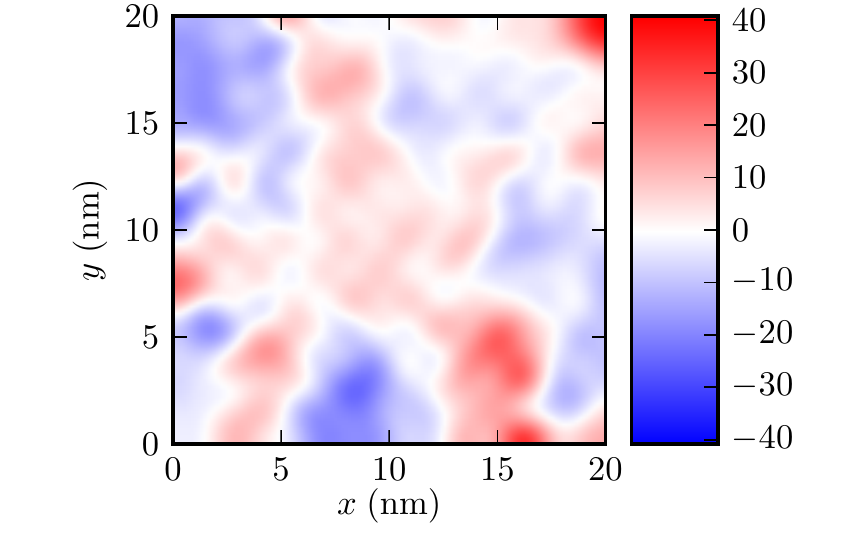} &
\includegraphics [width=0.33\linewidth]{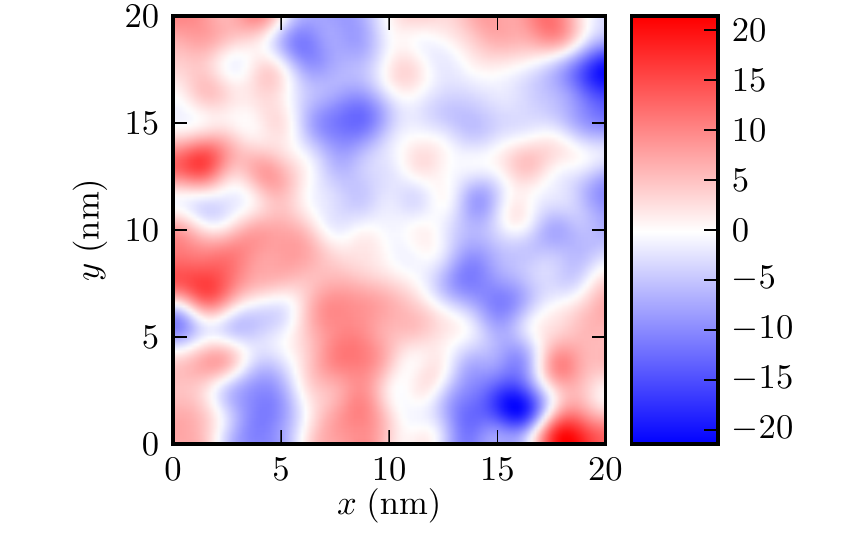} &
\includegraphics [width=0.33\linewidth]{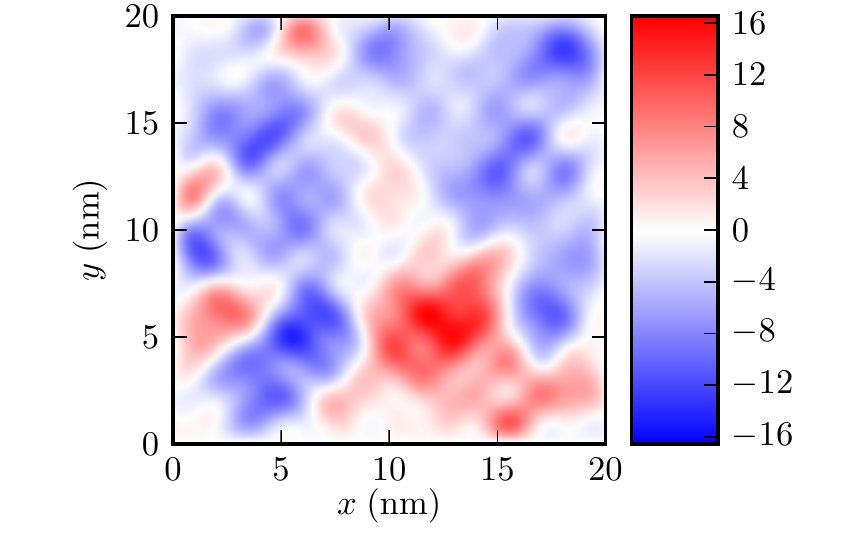}
\end{tabular}
\caption{(Color online)  Left panel: color plot of the scalar potential $V_1({\bm r})$ (in units of meV) calculated using Eq.~(\ref{eq:potentialsfromripples}) with $g_1 = 3~{\rm eV}$. Central panel: the ${\hat {\bm x}}$-component $A_x({\bm r})$ of the vector potential (in units of meV) calculated using Eq.~(\ref{eq:potentialsfromripples}). Right panel: same as in the central panel but for the ${\hat {\bm y}}$-component $A_y({\bm r})$ of the vector potential.\label{fig:two}}
\end{center}
\end{figure*}

The real-space scalar potential $V_1({\bm r})$ and the two components of the vector potential ${\bm A}({\bm r})$ calculated from Eq.~(\ref{eq:potentialsfromripples}) for $g_1 = 3~{\rm eV}$ and for the sample in Fig.~\ref{fig:one} have been reported in Fig.~\ref{fig:two}.
Since the experimental sample does not respect periodic boundary conditions (which are used in the numerical calculations below) we actually work with a $40~{\rm nm} \times 40~{\rm nm}$
sample which has been obtained by suitably replicating the original one~\cite{replicas}. All numerical results shown in this Rapid Communication refer to the experimentally-relevant portion of the simulation box.

{\it Self-consistent Kohn-Sham-Dirac theory of the induced carrier density. ---} The external scalar $V_1({\bm r})$ and vector ${\bm A}({\bm r})$ potentials plotted in Fig.~\ref{fig:two} and calculated from Eq.~(\ref{eq:potentialsfromripples}) are responsible for carrier-density inhomogeneities, which can be quantified by the deviation $\delta n({\bm r})$ of the local density $n({\bm r})$ from the ``background" value $n_0 =2\eta /{\cal A}_0+ {\bar n}_{\rm c}$. Here $2/{\cal A}_0$ is the density of a neutral graphene sheet, ${\cal A}_0=3\sqrt{3} a^2_0/2 \approx 0.052~{\rm nm}^2$ being the area of the unit cell in the honeycomb lattice, and ${\bar n}_{\rm c}$ is the spatially-averaged carrier density, which can be positive or negative and controlled by gate voltages. The dimensionless parameter $\eta\ll 1$ controls the fraction of $\pi$-band electrons that are described by the massless Dirac fermion model~\cite{reviews}. In the numerical calculations below $\eta \approx 0.1$.

Since $V_1({\bm r})$ and ${\bm A}({\bm r})$ change smoothly over many lattice constants, the induced density $\delta n({\bm r})$ can be calculated~\cite{polini_prb_2008,gibertini_prb_2010} by solving a
single-valley (and single-spin) Kohn-Sham-Dirac (KSD) equation for a two-component spinor $\Phi_\lambda({\bm r})= (\varphi^{(A)}_\lambda({\bm r}),\varphi^{(B)}_\lambda({\bm r}))^{\rm T}$:
\begin{equation}\label{eq:ksd}
\left\{{\bm \sigma}\cdot [v{\bm p} + {\bm A}({\bm r})] + \openone_{\sigma} V_{\rm K S}({\bm r}) \right\}\Phi_\lambda({\bm r})=\varepsilon_{\lambda}\Phi_\lambda({\bm r})~.
\end{equation}
Here ${\bm \sigma}$ is a 2D vector constructed with the $2 \times 2$ Pauli matrices $\sigma_1$ and $\sigma_2$ acting in sublattice-pseudospin space, $v = 3 \gamma_0 a_0/(2 \hbar) \approx 10^{6}~{\rm m}/{\rm s}$ is the bare Fermi velocity, ${\bm p}=-i\hbar \nabla_{\bm r}$, $\openone_{\sigma}$ is the $2 \times 2$ identity matrix in pseudospin space, and the Kohn-Sham potential,
\begin{equation}\label{eq:kspot}
V_{\rm KS}({\bm r}) = V_1({\bm r}) +  V_{\rm H}({\bm r}) + V_{\rm xc}({\bm r})~,
\end{equation}
is the sum of the external scalar potential $V_1({\bm r})$, the Hartree potential, and the scalar exchange-correlation (xc) potential.

The (classical electrostatic) Hartree potential is given by
\begin{equation}\label{eq:hartree}
 V_{\rm H}({\bm r})=\int d^2{\bm r}'\frac{e^2}{\epsilon|{\bm r}-{\bm r}'|} \; \delta n({\bm r}')~,
\end{equation}
where $\epsilon  = (\epsilon_{\rm vac} + \epsilon_{\rm sub})/2$ is an average dielectric constant,
$\epsilon_{\rm vac}$ ($\epsilon_{\rm sub}$) being the dielectric constant of the medium above (below) the graphene flake.
For example $\epsilon \approx 2.5$ for graphene placed on ${\rm SiO}_2$ (the other side being exposed to air), while
$\epsilon \approx 1$ for suspended graphene.

The third term in $V_{\rm KS}({\bm r})$, $V_{\rm xc}({\bm r})$, is the xc potential,
a functional of the ground-state density, which is known only approximately.
Following Refs.~\onlinecite{polini_prb_2008} and~\onlinecite{gibertini_prb_2010} we employ the local-density approximation (LDA),
\begin{eqnarray}\label{eq:LDA}
V_{\rm xc}({\bm r}) \stackrel{\rm LDA}{=} \left.\frac{d [n \delta\varepsilon_{\rm xc}(n)]}{dn}\right|_{n \to {\bar n}_{\rm c} + \delta n({\bm r})}~,
\end{eqnarray}
where $\delta\varepsilon_{\rm xc}(n)$ is the excess xc energy of a homogeneous 2D liquid of massless Dirac fermions with carrier density $n$~\cite{polini_prb_2008,barlas_prl_2007}.

The ground-state density $n({\bm r})$ is obtained as a sum over the KSD spinors $\Phi_\lambda({\bm r})$:
\begin{eqnarray}\label{eq:density}
n({\bm r}) = N_{\rm f} \sum_{\lambda}[|\varphi^{(A)}_{\lambda}({\bm r})|^2+|\varphi^{(B)}_{\lambda}({\bm r})|^2]
n_{\rm F}(\varepsilon_\lambda)~,
\end{eqnarray}
where the factor $N_{\rm f} = 4$ is due to valley and spin degeneracies and $n_{\rm F}(E)$ is the usual Fermi-Dirac thermal factor. Equation (\ref{eq:density}) is a self-consistent closure relationship for the KSD equation (\ref{eq:ksd}), since the Kohn-Sham potential $V_{\rm KS}({\bm r})$ is a functional of the ground-state density $n({\bm r})$.

Technical details on how to solve Eqs.~(\ref{eq:ksd})-(\ref{eq:density}) are discussed at great length in Refs.~\onlinecite{polini_prb_2008,gibertini_prb_2010}.

\begin{figure}
\begin{center}
\begin{tabular}{c}
\includegraphics[width=1.0\linewidth]{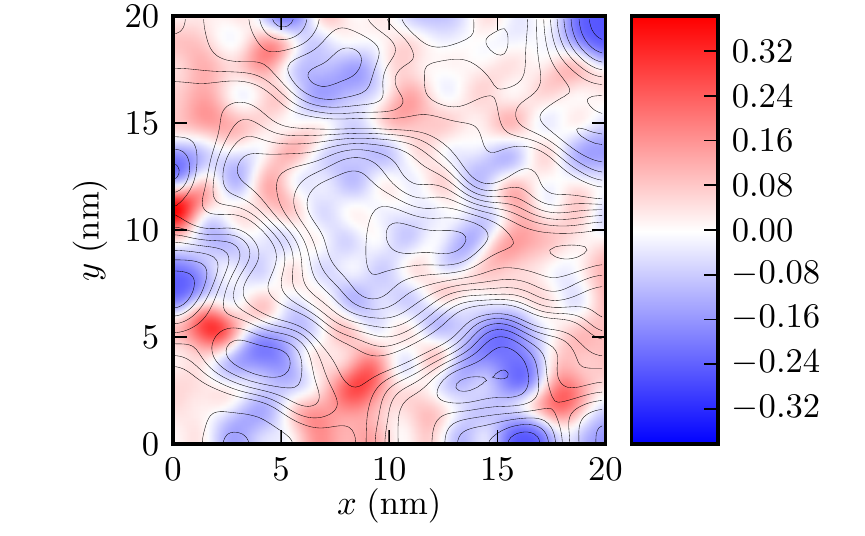}\\
\includegraphics[width=1.0\linewidth]{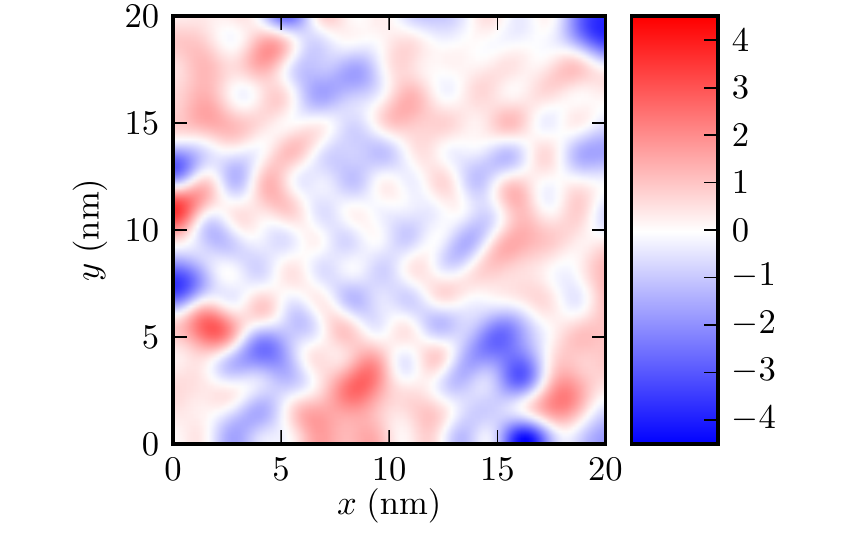}
\end{tabular}
\caption{(Color online) Top panel: Fully self-consistent  induced carrier-density profile $\delta n({\bm r})$
(in units of $10^{12}~{\rm cm}^{-2}$) in the corrugated graphene sheet shown in Fig.~\ref{fig:one}. The data reported in this figure have been obtained by setting $g_1 = 3~{\rm eV}$, $\alpha_{\rm ee} = 0.9$, and an average carrier density ${\bar n}_{\rm c} \approx 2.5\times 10^{11}~{\rm cm}^{-2}$. The thin solid lines are contour lines of the height map $h({\bm r})$. Note that there is no simple correspondence between topographic out-of-plane corrugations and carrier-density inhomogeneity. Bottom panel: same as in top panel but for $g_1 = 20~{\rm eV}$.\label{fig:three}}
\end{center}
\end{figure}

{\it Numerical results and discussion. ---} The color coding in Fig.~\ref{fig:one} represents the spatial map of the calculated induced carrier density $\delta n({\bm r})$ for a value of the graphene's fine-structure constant $\alpha_{\rm ee} \equiv e^2/(\hbar v \epsilon) =0.9$ (a value commonly used value for a graphene sheet on a ${\rm SiO}_2$ substrate). We remind the reader that $\alpha_{\rm ee}$ has the physical meaning of a dimensionless coupling constant that determines the strength of electron-electron interactions~\cite{reviews}. A 2D color plot of $\delta n({\bm r})$ is also reported in Fig.~\ref{fig:three} for the sake of clarity. In this figure we have presented predictions for $g_1= 3~{\rm eV}$ (as in Fig.~\ref{fig:one}) but also for $g_1= 20~{\rm eV}$. We clearly see that the carrier density profile $\delta n({\bm r})$ breaks into electron-hole puddles with extensions ranging from a few nanometers to the sample size.  Changing the value of $g_1$ from $3~{\rm eV}$ to $20~{\rm eV}$ leads merely to a change in the amplitude of carrier-density fluctuations but not in the spatial pattern of electron-hole puddles. Since the KSD theory includes screening due to $\pi$ electrons, we tend to think that one should use the unscreened value $g_1 \approx 20~{\rm eV}$ to avoid a double-counting of screening~\cite{g-one-discussion}. Note also the well-defined regions of zero induced density, an effect that can be traced back to the anomalous behavior of the xc potential in systems of massless Dirac fermions~\cite{polini_prb_2008,gibertini_prb_2010}.

A more quantitative analysis than that reported in Fig.~\ref{fig:one} 
of the degree of correlation between topographic out-of-plane corrugations and carrier-density inhomogeneities 
is shown in Fig.~\ref{fig:three}. Here we plot together with $\delta n({\bm r})$ contour lines of the height map $h({\bm r})$. 
From this figure one infers marginal correlations between topography and electron-hole puddles, as already noticed in Refs.~\onlinecite{gibertini_prb_2010,partoviazar_prb_2011} for simulated ripples. 
More mathematically, the real-space scalar and vector potentials that one derives from Eq.~(\ref{eq:potentialsfromripples}) are complicated
functionals~\cite{gazit_prb_2009} of the tensor field $f_{ij}({\bm r})$, {\it i.e.} of the height-fluctuation map $h({\bm r})$. For example, the scalar potential, 
is given ({\it modulo} a constant) by the following highly non-local expression
\begin{equation}\label{eq:scalar-realspace}
V_1({\bm r}) = \frac{g_1}{2\pi} \frac{\mu}{\lambda + 2 \mu}~\int d^2{\bm r}' \log{(|{\bm r} - {\bm r}'|)} {\cal F}({\bm r}')~,
\end{equation}
where ${\cal F}({\bm r}) = \sum_{i,j}(\delta_{ij} \nabla^2_{\bm r} - \partial_i\partial_j)f_{ij}({\bm r})$ is the Fourier transform of ${\cal F}({\bm q})$. As a consequence, carrier-density inhomogeneities are not correlated in a trivial fashion with the height map $h({\bm r})$. This is most transparent within linear-response theory in the random phase approximation~\cite{polini_prb_2008}. In this limit it is possible to show that
the induced density in response to $V_1({\bm r})$ for a neutral-on-average graphene sheet is given by
\begin{equation}\label{eq:LRT}
\delta n({\bm r}) = \int d^2 {\bm r}' \frac{q^2_{\rm eff}}{|{\bm r} - {\bm r}'|} {\cal F}({\bm r}')~,
\end{equation}
where the coupling constant $q^2_{\rm eff}$ (with physical dimensions of inverse length) is given by
\begin{equation}\label{eq:effectivecharge}
q^2_{\rm eff} = \frac{N_{\rm f}}{32\pi \hbar v}\frac{\mu}{\lambda + 2 \mu} \frac{g_1}{\displaystyle 1 +\frac{\pi}{8}N_{\rm f} \alpha_{\rm ee}}~.
\end{equation}
In deriving Eq.~(\ref{eq:LRT}) we have used that the static density-density response function of 2D non-interacting Dirac fermions is $\chi_0(q) = - N_{\rm f} q/(16 \hbar v)$. Eqs.~(\ref{eq:LRT})-(\ref{eq:effectivecharge}) capture qualitatively  the main features of the numerical solution of the self-consistent KSD equation even though they miss some important non-linear effects. Note i) the intriguing formal analogy between Eq.~(\ref{eq:LRT}) and the expression for the classical electrostatic potential in Eq.~(\ref{eq:hartree}) and ii) that the coupling constant $q^2_{\rm eff}$ depends on the screened value of $g_1$, ${\widetilde g}_1 = g_1/(1+\pi N_{\rm f}\alpha_{\rm ee}/8)$. Moreover, according to Eq.~(\ref{eq:LRT}), a reduction of the typical height fluctuations $h$ by an order of magnitude, implies a suppression of the amplitude $\delta n$ of density inhomogeneities by two orders of magnitude, in agreement with recent observations for graphene on h-BN~\cite{ xue_naturemater_2011,decker_nanolett_2011}.

In summary, we have shown that in a real sample corrugation-induced scalar and vector potentials alone can in principle lead to carrier-density inhomogeneities with length scales that are larger than the spatial resolution of current scanning tunneling microscopes~\cite{STMgraphene}. A serious comparison between experimentally-reconstructed carrier-density profiles and our theoretical predictions may lead in a near future to achieve a better understanding of the main mechanism leading to electron-hole puddles and limiting the mobility of unsuspended samples. While this paper focusses on graphene sheets on quartz, we believe that it would be very interesting to carry out extensive comparisons between our theory and experimental data for graphene flakes on h-BN~\cite{ xue_naturemater_2011,decker_nanolett_2011}.

{\it Acknowledgements. ---} Work in Pisa was supported by the Italian Ministry of Education, University, and Research (MIUR) through the program ``FIRB - Futuro in Ricerca 2010" (project title ``PLASMOGRAPH: plasmons and terahertz devices in graphene").  F.G. gratefully acknowledges MICINN (Spain) through grants FIS2008-00124 and CONSOLIDER CSD2007-00010. M.I.K. acknowledges financial support by the 
Stichting voor Fundamenteel Onderzoek der Materie (FOM) (The Netherlands). 
We gratefully acknowledge Viktor Geringer for sending us experimental data on the height fluctuations of a graphene sheet on ${\rm SiO}_2$.


\begin{thebibliography}{77}
\bibitem{replicas}
	The $40~{\rm nm} \times 40~{\rm nm}$ sample with periodic boundary conditions has been constructed by juxtaposing the original sample with three samples obtained from 
	the original one by reflection through a horizontal axis, by reflection through a vertical axis, and by spatial inversion, respectively.
\end{thebibliography}
\end{document}